\documentclass[graybox]{svmult}
\usepackage{graphicx}
\usepackage{amsmath}
\usepackage{subcaption}
\usepackage{comment}
\usepackage{makeidx}         
\usepackage{graphicx}        
\usepackage{multicol}        
\usepackage[bottom]{footmisc}
\makeindex
\begin{document}
\title*{Expressing Trust with Temporal Frequency of User Interaction in Online Communities}


\author{Ekaterina Yashkina, Arseny Pinigin, JooYoung Lee, Manuel Mazzara, Akinlolu Solomon Adekotujo, Adam Zubair, Luca Longo}
\institute{Ekaterina Yashkina, Arseny Pinigin, JooYoung Lee and Manuel Mazzara \at Innopolis University, Russia \email{\{e.yashkina, a.pinigin, j.lee, m.mazzara\}@innopolis.ru}
\and Akinlolu Solomon Adekotujo and Adam Zubair \at Lagos State University, Nigeria
 \email{z.folohunsho@innopolis.university,  adekotujoakinlolu@gmail.com}
 \and Luca Longo$^*$ \at School of Computer Science, Technological University Dublin, Republic of Ireland. \email{luca.longo@dit.ie}
 }

\authorrunning{Ekaterina Yashkina et al.}

\maketitle

\thispagestyle{plain}
\pagestyle{empty}

\abstract{
Reputation systems concern soft security dynamics in diverse areas. Trust dynamics in a reputation system should be stable and adaptable at the same time to serve the purpose. Many reputation mechanisms have been proposed and tested over time. However, the main drawback of reputation management is that users need to share private information to gain trust in a system such as phone numbers, reviews, and  ratings. Recently, a novel model that tries to overcome this issue was  presented: the Dynamic Interaction-based Reputation Model (DIBRM). This approach to trust considers only implicit information automatically deduced from the interactions of users within an online community. In this primary research study, the {\it Reddit} and {\it MathOverflow} online social communities have been selected for testing DIBRM. Results show how this novel approach to trust can mimic behaviors of the selected reputation systems, namely {\it Reddit} and {\it MathOverflow}, only with temporal information.}

\section{Introduction}
Numerous approaches haven been proposed to capture trust level of users in a system \cite{massa2005controversial,massa2007survey,hsu2011exploring, gal2011identifying,dondio2014computing} and their propagation \cite{ruan2014exploring}. Among others, reputation mechanisms are widely adopted by online social media to enable secure and effective interactions among users~\cite{tausczik2014collaborative,Matthews:2016:GMN:2872518.2890592, zhang2011influence, bharadwaj2009fuzzy}.
Trust dynamics in a reputation system should be stable and adaptable at the same time to serve the purpose \cite{abdul2000supporting,luca2009enabling}. Typically, reputation mechanisms consider factors such as feedback on previous interaction, frequencies of interactions, the presence of a user in an underlying system or community~\cite{tormo2015dynamic,sherchan2013survey}. 
However, the main drawback of reputation management is that users need to share private information to gain trust in a system such as phone
numbers, reviews, ratings, and other pieces of information. This might impact the system itself because some users are not willing to share their sensitive information~\cite{hoffman2009survey}. 
A recent work  proposed a new reputation mechanism that only considers the frequency of users interactions in order to compute their reputation value~\cite{DBLP:conf/aina/MelnikovLRML18}. This approach, named the Dynamic Interaction-Based Reputation Model (DIBRM), was preliminary tested with data from the StackOverflow online community. Findings suggested that by only considering information related to the frequency of the interactions of users, DIBRM could closely mimic the original reputation scores proposed by StackOverflow. Therefore, DIBRM can have a competitive advantage over other reputation mechanisms since it is fully automated without require intervention and explicit  information from users \cite{LongoBD09}.\\

The goal of this research is to further test and validate the DIBRM mechanism when applied to other  social networks, namely {\it Reddit} and {\it MathOverflow}. The remained of the paper is organized as follows.
Section \ref{relatedWork} introduces related work on reputation and trust  emphasizing some limitations. Section \ref{DesignMethodology} builds on such limitations and describe a novel primary research experiment employing the aforementioned DIBRM mechanism.
Results are presented in Section~\ref{Results} and discussed in Section~\ref{discussion}. Eventually, Section~\ref{conclusions} summarizes the contribution to the body of knowledge and recommend future research.

\section{Related Work}\label{relatedWork} 
Reputation systems are essential in many  application domains such as online communities~\cite{josang2007survey}. They exist as a form of soft security measure to overcome the issues generated by those individuals who legitimately cheat the system. Different types of reputation mechanisms have been introduced in the last years \cite{sabater2005review}. These mechanisms usually sees an online community as a directed graph whereby nodes either represent a user or a piece of information, and an arrow models the interaction between two nodes~\cite{buchegger2008reputation}.
From a graph, topological information can be extracted such as the position of each node or the interactions between nodes~\cite{Lee:2013:MRP:2492517.2492663}. Here, a reputation  model, called ReMSA (Reputation Management for Social Agents) is built upon users' feedback and voting as well as the time decaying necessary to update the reputation of users every time a new interaction is added to the graph. The voting mechanism is recursive and a node can collect pieces of feedback, about the target node, from remote nodes that are not directly connected to it. 
Brown et al.~\cite{brown2017reputation} recently proposed a reputation system for Non-Player Characters (NPC) in online games. The authors focused on how the reputation of users can be realistically propagated by considering for instance, their communication, relationships and physical constraints.
It is interesting to see that the resulting communities, based on interactions of players, also match the environments (cities or communities) in online games.
Although many reputation mechanisms are means for controlling systems and therefore centralized, some distributed reputation algorithm exists as well. For example, a decentralized reputation mechanism was presented in~\cite{yu2002evidential}. Here, each user in a community is involved in the subjective assessment of the reputation of other users. This is useful when there is no implicit reputation scheme as it always occurs in peer-to-peer systems. However, this mechanism is explicit and its scalability represents the main drawback.
In many social networks or other online communities containing information in the form of news articles, technical contents or comments from users, an explicit or implicit mechanism for rating contents or users does not exist \cite{LongoBD09}. 
To overcome this limitation, a body of literature is devoted to assign reputation values to entities (users of information) automatically \cite{luca2009information,longo2010enhancing,dondio2011trust}.
For example, the algorithm presented in~\cite{5992663} can automatically extract the reputation of a user from published textual content using natural language analysis. \\

The work by Longo et al.~\cite{longo2007temporal} tested the hypothesis that temporal-based factors, such as the activity, frequency, regularity of interactions of an entity and its presence in a community, can be used as evidence of its trustworthiness. This hypothesis was tested considering Wikipedia and its content (12000 users, 94000 articles). The authors successfully demonstrated how their algorithm could identify top Wikipedia contributors similarly to the explicit `barnstars' mechanism~\footnote{https://en.wikipedia.org/wiki/Wikipedia:Barnstars} employed by Wikipedia.
The main drawback of using this approach is the computational time required to quantify each temporal factors for each users which exponentially increases in the number of users and their interactions.
To overcome this limitation, a simpler mechanism was introduced by Melinkov et al.~\cite{DBLP:conf/aina/MelnikovLRML18}. Here, the frequency of interactions among users is uniquely considered to compute their reputation. This approach was tested by employing information collected from {\it StackOverflow}~\footnote{https://stackoverflow.com/}. Findings show how this mechanism can closely approximate the reputation scores inferred by the {\it StackOverflow} mechanism. 

\section{Design and methodology}\label{DesignMethodology}
A primary research study has been conducted to investigate the capability of the Dynamic Interaction-based Reputation Model (DIBRM)~\cite{DBLP:conf/aina/MelnikovLRML18} for assessing the reputation of users in online communities. In details, two social online communities, namely  {\it Reddit}~\footnote{https://www.reddit.com/} and {\it MathOverflow}~\footnote{https://mathoverflow.net/}, have been selected for  such an investigation. The remainder of this section is devoted to the formal description of the DIBRM model, its application to the two selected online platforms and the strategy employed for its evaluation.

\subsection{The Dynamic Interaction-based Reputation Model}
Interactions $I_n$ in DIBRM are modeled as:
$$I_n=I_{b_n}+I_{c_n}$$
where $n \in 0\ldots N$ is the index of the interaction and $N$ is the total number of interactions of a user. $I_n$ contains a time stamp indicating that moment in time when an interaction took place and a value that describes the contribution to the reputation. These values can be enumerated by the associated time stamp to form a chain of historical user's activity.
Interactions have different effects on the trust of a user. Each interaction has a basic value $I_{b_n}$. For example, in {\it MathOverflow}, an interaction could be `asking', `answering', and `editing'. Each interaction has a fixed contribution to the user's reputation value and we consider it as a basic value. 
Depending on the state of communication between a user and the system characterized by activity and frequency, an interaction can be perceived differently. $I_{c_n}$ captures the cumulative part of an interaction.
It is defined as:
$$I_{c_n}=I_{b_n}*\alpha*(1-\frac{1}{A_{n}+1})$$
where $\alpha$ is the weight of the cumulative part. This indicates how much $I_{c_n}$ can grow (if $\alpha = 1$ then $I_{c_n} \in 0\ldots I_{b_n}$). $A_{n}$ is the number of  activities for a user. 
Online social communities have different contexts and features that can affect the properties of the trustworthiness of its entities. One of these properties is the frequency of the  communication of an user which is defined as the period of time between the last two $I_n$ activities performed. DIBRM models this property as $t_a$. As an example, $t_a$ for Wikipedia, on one hand, can be set to one week, time period within which a user can create or edit some article. On the other hand, for StackOverflow, it can be set to one day, indicating the interval of time within which a user can answer to a certain online post (question in this context).  The number of periods between the 2 last interactions ($t_n$ and $t_{n-1}$) of a user can be formalised as:
$$\Delta_n=\bigg[\frac{t_n - t_{n-1}}{t_a}\bigg]$$
If the difference between $t_n$ and $t_{n-1}$ is less than $t_a$, then the amount of activity periods will increase by one. It means that a user  interacts frequently within the online community. 
Eventually, the final degree of trust for a certain user can be formalized as: 
$$T_n = T_{n-1}*\beta^{\Delta_n} + I_n, \beta \in [0,1]$$
where $\beta$ is the forgetting factor, the rate reputation decays,that is chosen by each system individually. The more $\beta$ approximates to $1$, then more the trust value of a user decreases.
Another parameter considered by DIBRM is the sum of the previous trust values computed for a given user. We refer to this parameter as {\it historical} reputation. In the evaluation section, we use this parameter to compare DIBRM against other reputation models because this value accumulates historical information about a user’s reputation. 
In other words, if a user currently has a low reputation but was very active in the past, having interacted within the community multiple times, its {\it historical} reputation can still be high.

\subsection{Reddit and MathOverflow}
{\it Reddit} is an American social news aggregator, web-content rating and discussion platform~\cite{singer2014evolution}. Registered users can submit entries to it as texts, images and links. These entries can be up or down voted by other users. 
 Since the official reputation algorithm employed by {\it Reddit} is not publicly available, although  speculations exist ~\cite{salihefendic2010reddit}, its detailed description cannot be provided. 
However, what is known is that the time of submission of an entry, in the {\it Reddit} online platform, is an important element for rankings posts. In fact, those posts with a similar number of `upvotes' and `downvotes'  are ranked lower compared to other posts with `upvotes' only. 
{\it Reddit} also sorts {\it comments} in a different manner than {\it posts}, and it does not consider temporal information, displaying the most popular first~\cite{salihefendic2010reddit}. 
Additionally, two other types of information are computed: `post karma' and `comment karma' which is how much a user has contributed to the community.
The score assigned to a particular post depends mostly on the date of submission. Newer publications have higher scores when compared to older publications even if the amount of votes is the same. \\

{\it MathOverflow} is a question and answer online platform for mathematicians. It allows its users to ask questions, submit answers, rate them and collect points for their activities. A user who asked a question can only accept one answer to be rewarded with points even if there are several correct answers. However, users can `upvote' multiple answers. According to {\it MathOverflow}, a user's reputation  depends on three activities: asking, answering and editing. Restrictions exist to avoid malicious behaviors. Accepting own answer does not increase own reputation. Deleted posts do not affect reputation and if a user reverses a vote, so the corresponding reputation loss/gain.

\section{Evaluation}
To measure the efficiency of DIBRM, a comparative analysis of its inferences, in form of a ranking of its users by their trustworthiness, is compared against the rankings of users produced by the \emph{Reddit} and \emph{MathOverflow} models.
In detail, an evaluation metric is defined:
$$\mu_D= 1 - \frac{1}{N^2}*\sum^{N}_{i=1}{\bigg(\frac{1}{D}*\sum^{D}_{j=1}{|R_{Re_{ij}} - R_{D_{ij}}|}\bigg)}$$
where $N$ is the number of users, $D$ the number of days between first and last dates, $R_{Re_{ij}}$ is the {\it Reddit} reputation value of user $i$ on day $j$ (analogously use $R_{Ma_{ij}}$ for {\it MathOverflow}) and $R_{D_{ij}}$ is the DIBRM reputation value of user $i$ on day $j$. $|R_{Re_{ij}} - R_{D_{ij}}|$ is the absolute difference between rating places of individual $i$ on particular day $j$. This value shows how close DIBRM rating is to {\it Reddit} and {\it MathOverflow}. Then the average difference of ratings for user $i$ $\frac{1}{D}*\sum^{D}_{j=1}{|R_{Re_{ij}} - R_{D_{ij}}|}$ (also for $R_{Ma_{ij}}$) in all-days period is calculated. It allows to avoid focusing on one estimation and analyse the general behavior of the model. After that, the average difference of all users is estimated. The last step focuses on subtracting the average difference from 1, which is then divided by the number of rating places $N$. This result gives information about how much the DIBRM rating model is close to {\it Reddit} and {\it MathOverflow}.

A second approach is to measure the ratings of users via an historical reputation value. The formula  remains the same but instead of $R_{D_{ij}}$ (reputation rating place of user $i$ on day $j$) $R_{H_{ij}}$ (historical reputation rating) is used.
$$\mu_H= 1 - \frac{1}{N^2}*\sum^{N}_{i=1}{(\frac{1}{D}*\sum^{D}_{j=1}{|R_{Re_{ij}} - R_{H_{ij}}|})}$$
Moreover, an error metric should be estimated to have a  clear picture of DIBRM. This is performed by calculating the standard deviation of the metric $\mu$. For reputation, it is $\sigma_D$, for historical reputation it is $\sigma_H$.

\subsection{Analysis of data}
Each system defines its own user reputation metrics. As previously mentioned, the procedure used by {\it Reddit} for computing user reputation is not public, therefore it is not possible to fully assess the similarities with the DIBRM model. {\it MathOverflow} system works similarly to {\it StackOverflow}. 
To evaluate  DIBRM, `votes', `posts', `comments' and users are all necessary pieces of information. 
The total number of users collected from {\it MathOverflow} over a period of one year is $4793$. {\it Reddit} data was  collected for a period of three months only. These amounts 
are deemed sufficient for comparative purposes. 
Data was converted into a csv-format for subsequent processing and tuples of the following form have been generated:
\begin{itemize}
    \item posts: $<$PostId, UserId, CreationDate, Vote$>$ 
    \item comments: $<$CommentId, UserId, CreationDate, PostId, Vote, ParentId$>$
\end{itemize}

\noindent In details:
\begin{enumerate}
    \item PostId, CommentId: identifier of an entity (integer)
    \item UserId: identifier of the author of an entity (integer)
    \item CreationDate: date of creation of an entity (date)
    \item Vote: amount of entity's approvals (integer)
    \item ParentId: entity's identifier  to which a comment was left to (integer)\\
\end{enumerate}

\noindent For the {\it Reddit} data, information on `posts' and `comments'  has to be divided since the {\it Reddit} rating (karma) consists of two parts: post karma and comments karma, these being computed differently. The overall rating is the sum of these two values. For {\it MathOverflow}, `posts' and `comments' can instead be stored together. This organisation using tuples allows to exploit both the {\it Reddit} and the {\it MathOverflow} reputation models.

\section{Experimental Results}\label{Results}
In this section, a comparison is done between the DIBRM inferences and those from {\it StackOverflow}, {\it Reddit} and {\it MathOverflow}.\\

\noindent \textbf{DIBRM}
The DIBRM model was tested with {\it StackOverflow} covering 4-year of data including $15,000$ users, $8,000,000$ posts and $33,000,000$ votes. The historical reputation approximation computed out of this data
 was $0.85$ with different values of $\alpha, \beta$ and $t_a$. This is  considered as a very good approximation. A possible interpretation of the results from tables ~\ref{table:4}-~\ref{table:2}, is given below.\\

\begin{table}[h]
\begin{minipage}[t]{0.58\textwidth}
\centering
\captionof{table}{Historical reputation \\scores for different $\beta$ (forgetting factor)}
\begin{tabular}{ | l | l | l | l | l |}
    \hline
    $\#$ & $t_a$ & $\beta$ & $\mu_H$ & $\sigma_H$ \\ \hline
    1 & 2 & 0.90 & 0,8803 & 0,0024  \\  \hline
    2 & 2 & 0.99 & 0,8805 & 0,0026  \\  \hline
    3 & 8 & 0.90 & 0,8808 & 0,0023  \\  \hline
    4 & 8 & 0.99 & 0,8808 & 0,0021  \\  \hline
    \end{tabular}
\label{table:4}
\end{minipage}
\begin{minipage}[t]{0.49\textwidth}
\centering
\captionof{table}{Reputation scores for \\ different $\alpha$ (cumulative factor)}
    \begin{tabular}{ | l | l | l | l | l |}
    \hline
    $\#$ & $\alpha$ & $\mu_D$ & $\sigma_D$ \\ \hline
    1 & 1 & 0,8313 & 0,0936  \\  \hline
    2 & 2 & 0,8441 & 0,0774  \\  \hline
    3 & 4 & 0,8426 & 0,0814 \\  \hline
    4 & 8 & 0,8515 & 0,0723  \\  \hline
    \end{tabular}
\label{table:5}
\end{minipage}
\begin{minipage}[t]{0.58\textwidth}
\centering
\captionof{table}{Reputation scores for \\  different $t_a$ (period)}
    \begin{tabular}{ | l | l | l | l |}
    \hline
    $\#$ & $t_a$ & $\mu_D$ & $\sigma_D$ \\ \hline
    1 & 1 & 0,8122 & 0,1100  \\  \hline
    2 & 2 & 0,8313 & 0,0936  \\  \hline
    3 & 4 & 0,8510 & 0,0744  \\  \hline
    4 & 8 & 0,8605 & 0,0604  \\  \hline
    \end{tabular}
\label{table:1}
\end{minipage}
\begin{minipage}[t]{0.49\textwidth}
\captionof{table}{Historical reputation \\scores for different $t_a$ (period)}
\centering    
\begin{tabular}{ | l | l | l | l |}
    \hline
    $\#$ & $t_a$ & $\mu_H$ & $\sigma_H$ \\ \hline
    1 & 1 & 0,8816 & 0,0128  \\  \hline
    2 & 2 & 0,8805 & 0,0026  \\  \hline
    3 & 4 & 0,8813 & 0,0086  \\  \hline
    4 & 8 & 0,8808 & 0,0021  \\  \hline
    \end{tabular}
\label{table:2}
\end{minipage}
\end{table}

\begin{itemize}
\item by increasing $t_a$ (period) makes trust scores slightly larger because users have a wider window for communications and interactions. However, if  the reputation change of two users over time is compared, the larger the $t_a$ gets, the larger the difference between two users is;
\item by increasing the forgetting factor ($\beta$) also the growth of the reputation scores increases because it gives more importance to past interactions (best results are achieved with $\beta$ = 0.99);
\item the cumulative factor ($\alpha$) shows how much the interaction value can grow as well how the  $I$ value can increase.
\end{itemize}
In summary, the DIBRM model succeeded in approximating the {\it StackOverflow} ratings with accuracy of $0.88$. The only disadvantage of DIBRM is parameter tuning.\\

\begin{figure}
    \centering
    \begin{subfigure}[b]{0.48\textwidth}
        \includegraphics[scale=0.19]{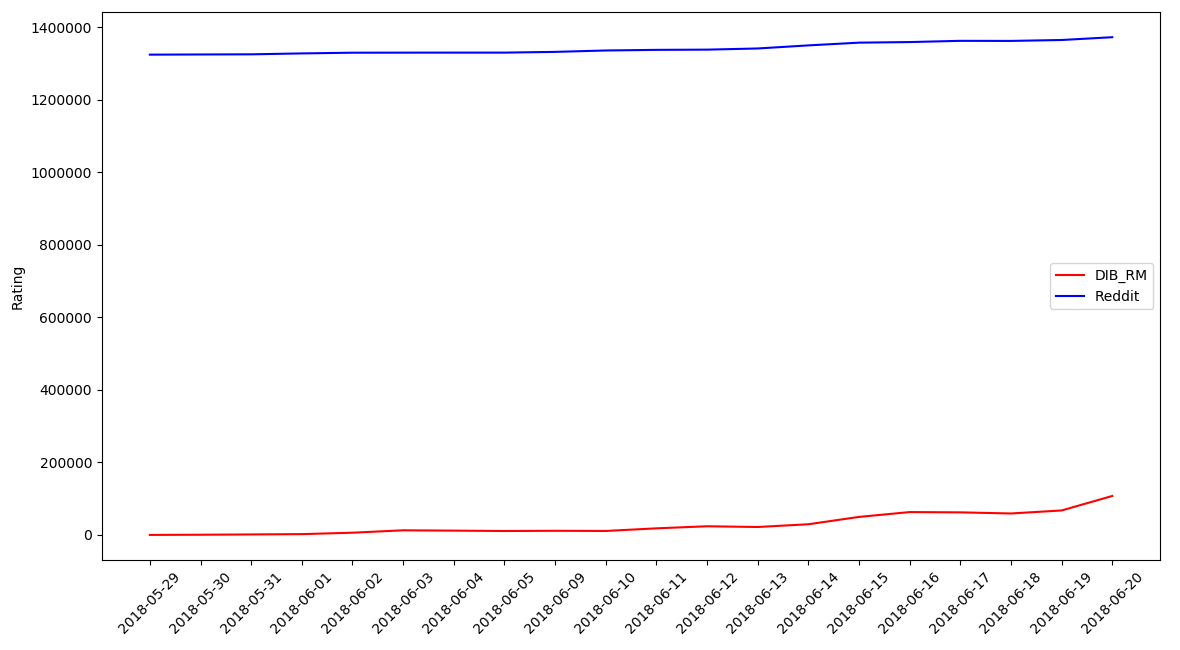}
        \caption{user {\it Meldroit1} (post)}
        \label{meldroit}
    \end{subfigure}
    ~ \hfill 
    \begin{subfigure}[b]{0.48\textwidth}
        \includegraphics[scale=0.192]{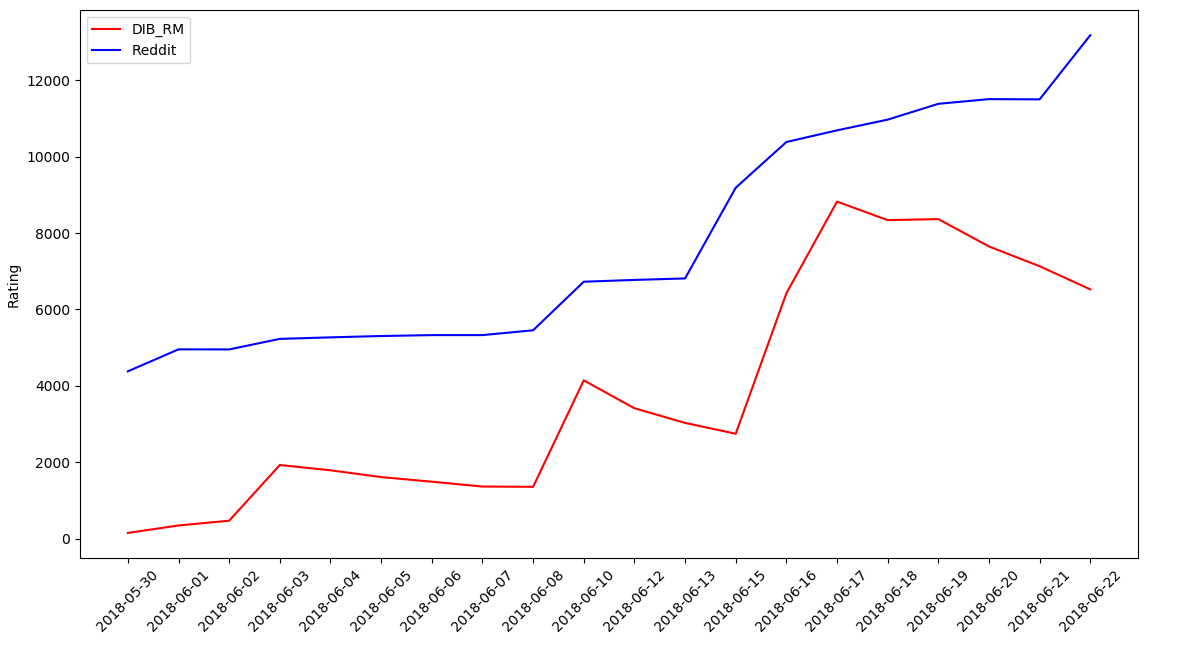}
        \caption{user {\it Gondala123} (post)}
\label{gondala}
    \end{subfigure}
    \begin{subfigure}[b]{0.48\textwidth}
        \includegraphics[scale=0.36]{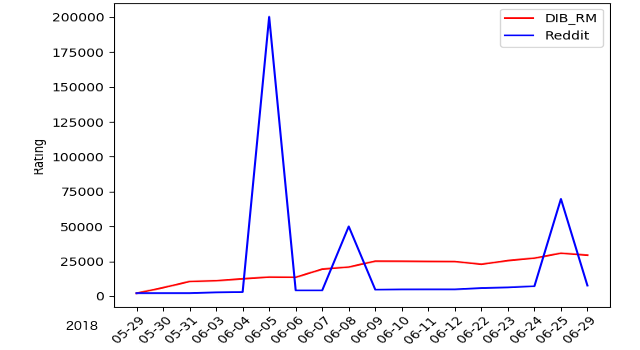}
        \caption{user {\it SergioVamos} (comment)}
\label{sergio}
    \end{subfigure}
    ~ \hfill
    \begin{subfigure}[b]{0.48\textwidth}
        \includegraphics[scale=0.193]{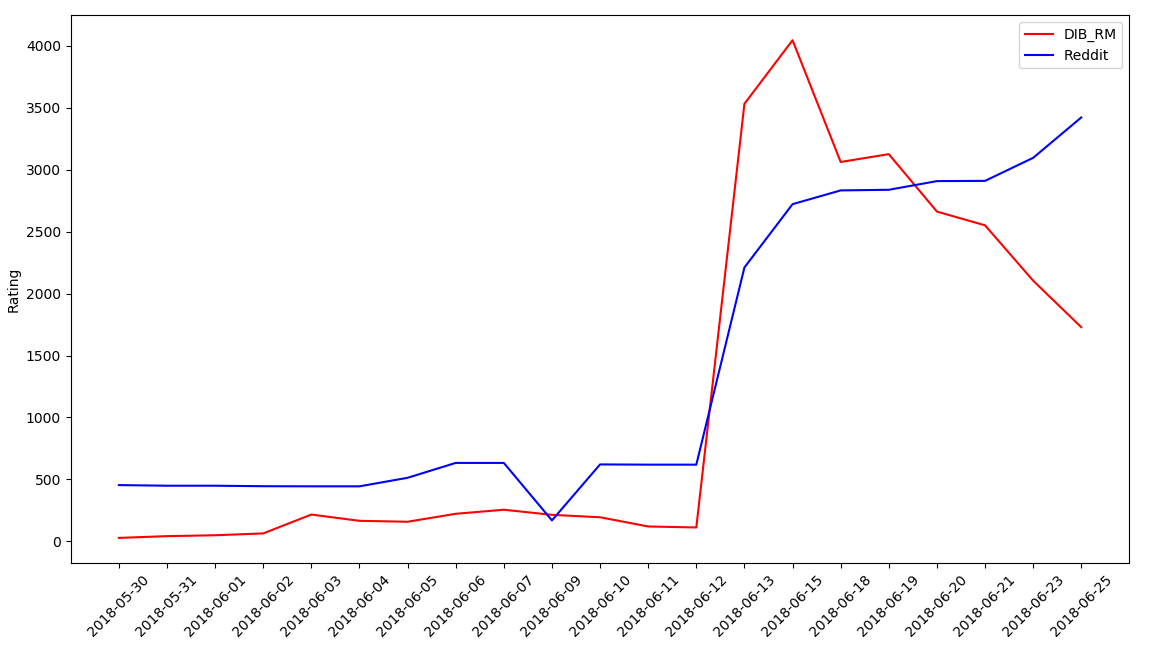}
        \caption{user {\it SurfingSilverSurfer} (comment)}
\label{surfing}
    \end{subfigure}
    \caption{Rating changes for different users for Reddit vs DIBRM.}
\label{eka:users}
\end{figure}

\noindent \textbf{Reddit}
From Figure~\ref{meldroit} it is possible to observe that the difference between DIBRM and {\it Reddit} ratings is significant (about 1.5 million). According to Table~\ref{table:eka}, this is similar to the average result for most of the users.

\begin{table}[h!]
\captionof{table}{DIBRM \& Reddit scores ($\alpha=1$, $\beta=0.9$, $t_a=1$ \text{day}).}
\begin{center}
    \begin{tabular}{ | c | r | r | r | r | r | r |}
    \hline
    Date & $\sharp$posts & $\sharp$post users & $\sharp$comments & $\sharp$comment users & post results & comment results\\ \hline
    $2018.05.27$ & 1000 & 800 & 14000 & 12000& -2305.657& -1600.453 \\  \hline
    $2018.05.30$ & 2500 & 2350 & 17000 & 15000& -172.952& -100.525 \\  \hline
    $2018.06.06$ & 4775 & 4000 & 33500 & 28000 & -21.163 & -0.304\\  \hline
    $2018.06.13$ & 8468 & 5129 & 57556 & 32008& -9.539 & 0.126\\  \hline
    $2018.06.20$ & 13367 & 7699 & 91225 & 47883& -5.278 & 0.455\\  \hline
    $2018.06.27$ & 17495 & 9737 & 121326 & 59870& -3.544 & 0.584\\  \hline
    $2018.07.02$ & 19282 & 10553 & 133349 & 64236& -3.047& 0.620\\  \hline
    \end{tabular}
\end{center}
\label{table:eka}
\end{table}

\noindent Figure~\ref{gondala} depicts an exception: this approximation is the overall best result, although the difference is still significant. Additionally, the fluctuation of the DIBRM ratings resembles the {\it Reddit} ratings, showing a good approximation.

\noindent The inferences with `comment' ratings are significantly better than those with `post' ratings. Figure~\ref{sergio} shows the approximation of DIBRM for `comment' ratings for the user {\it SergioVamos} which is very accurate, except for a few days. In Figure~\ref{surfing},  a typical trend of approximation is depicted with the two ratings close to each other with marginal differences.
The main reason why the accuracy is low  with the `post' ratings is likely due to the limited size of the dataset considered. When DIBRM was tested with the {\it StackOverflow} data, it covered roughly 3 years but, with the {\it Reddit} data, (manually collected in a short time frame), only 1 month  was considered.  From the results of table \ref{table:eka}, it can be observed that, with additional data, they can become more accurate (when the value is $1$, then the two results are similar, implying a perfect approximation). As a consequence, if the sample size is small, the difference between the actual and the DIBRM ratings is significant (Figure~\ref{meldroit}) as well as the ratio of the difference compared to the number of users. 
Additionally, when increasing $\alpha$ or $\beta$ individually, results dot not change much (about $0.0001$). However, if both  parameters are set to large values, then the results improve on a magnitude of $0.1$. The most suitable time period for {\it Reddit} was found to be $t_a=1$ day and with departure from this, inferences get worse.\\

\begin{figure}[h!]
\centering
\begin{subfigure}{0.99\textwidth}
\includegraphics[width=1\linewidth]{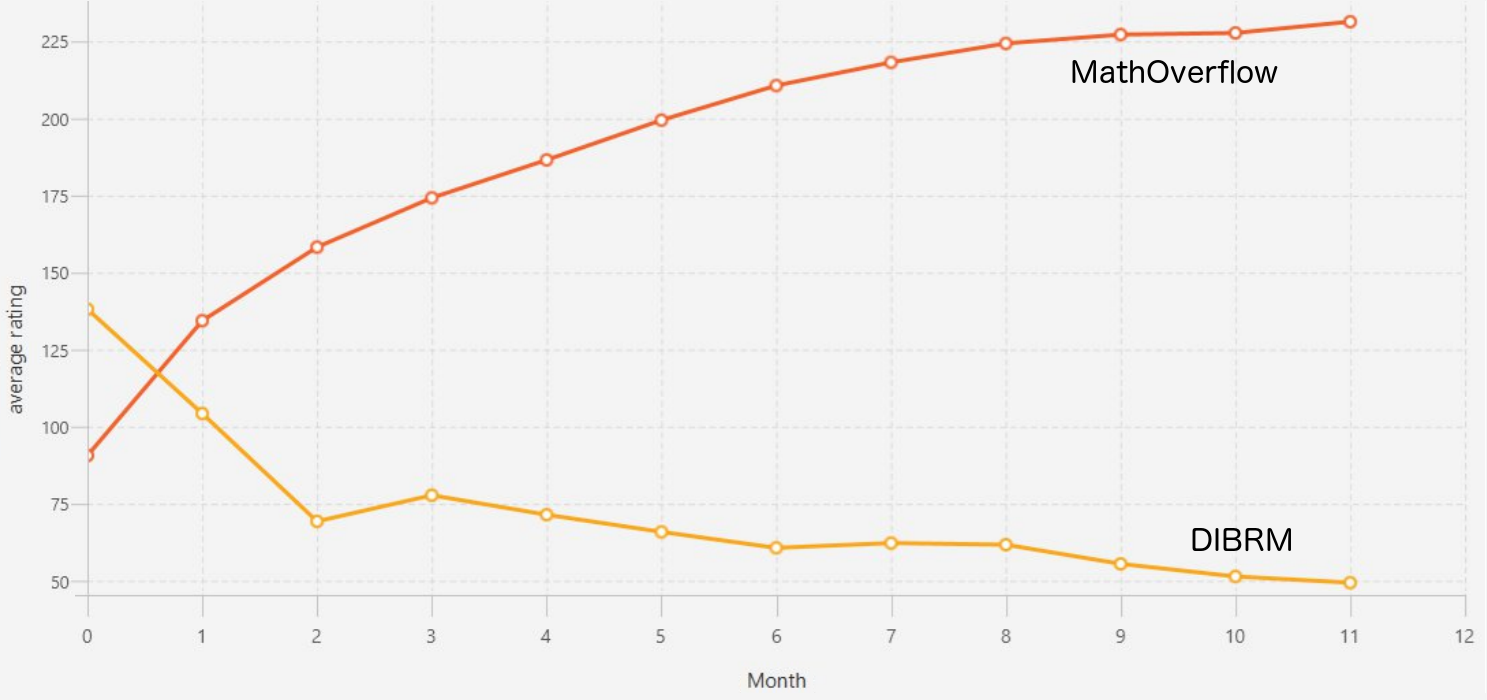} 
\caption{Average rating change over a month.}
\label{initial_ta}
\end{subfigure}
\begin{subfigure}{0.99\textwidth}
\includegraphics[width=1\linewidth]{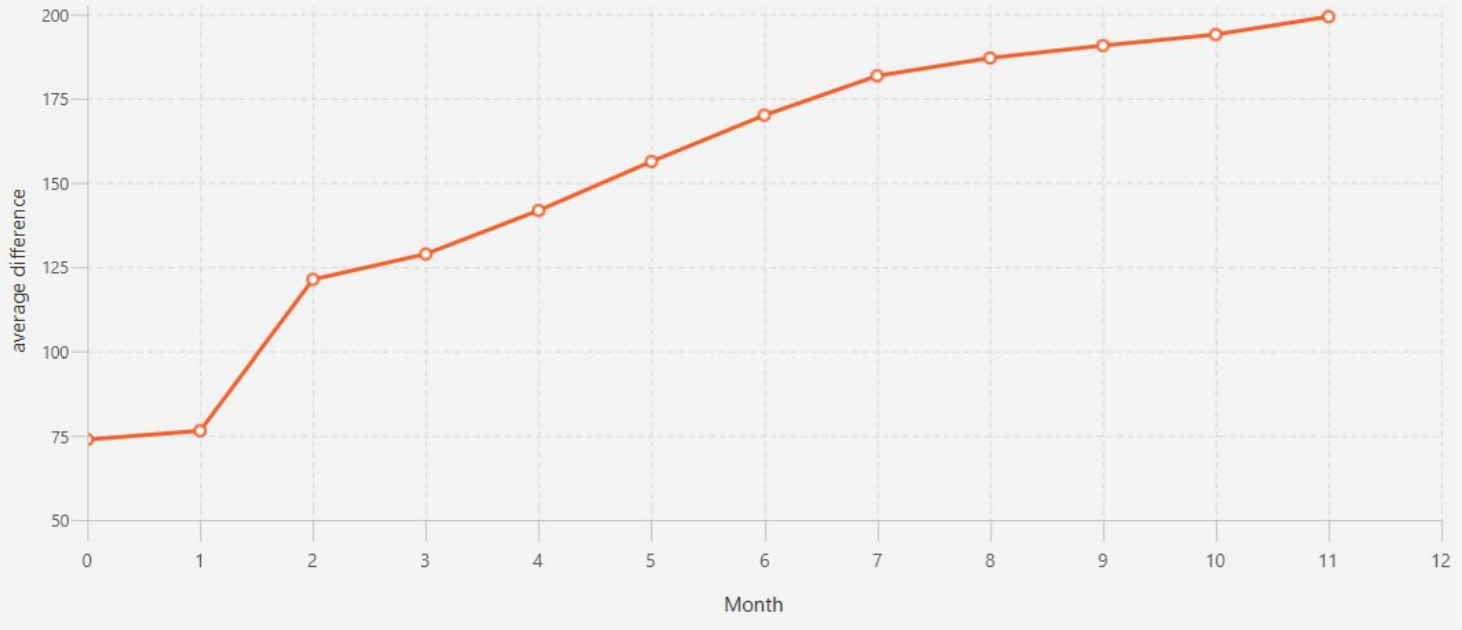} 
\caption{{Monthly difference between the average users reputation. The approximation was $\mu_D$ = 0.68 for this period.}}
\label{avediff}
\end{subfigure}
\caption{Dynamic reputation graphs for {\it MathOverflow} vs {\it DIBRM}.}
\end{figure}

\begin{figure}[h!]
\centering
\begin{subfigure}{0.89\textwidth}
\includegraphics[width=1\linewidth]{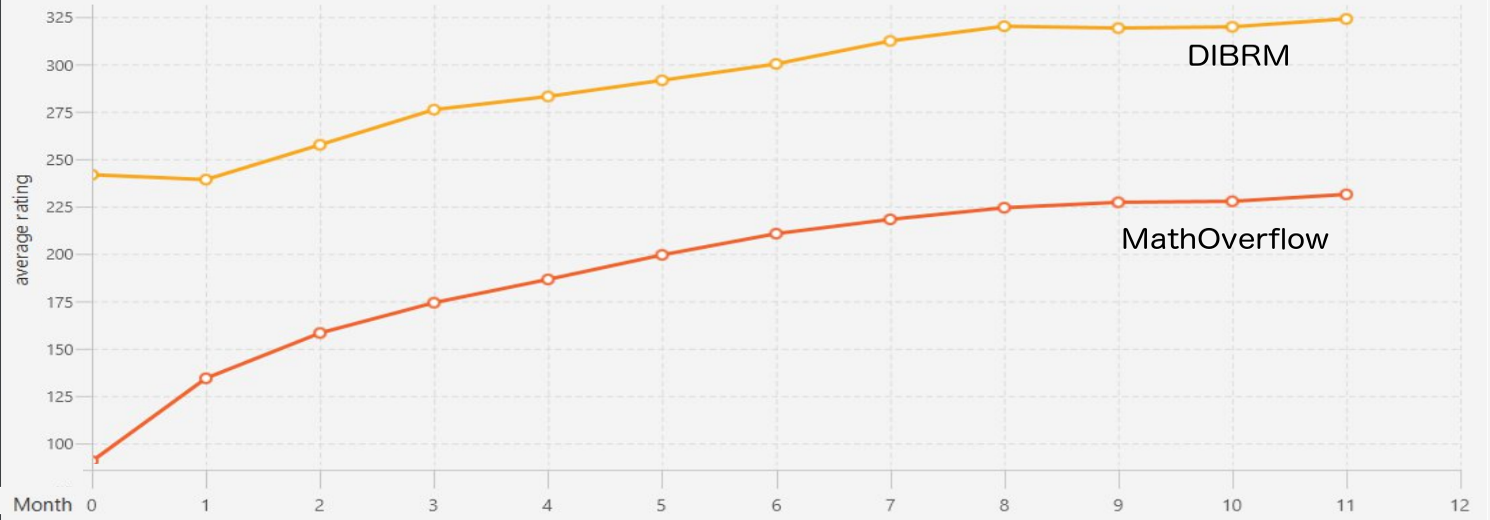} 
\caption{Average rating change, $t_a=8*10^4$ days.}
\label{ave}
\end{subfigure}
\begin{subfigure}{0.89\textwidth}
\includegraphics[width=1\linewidth]{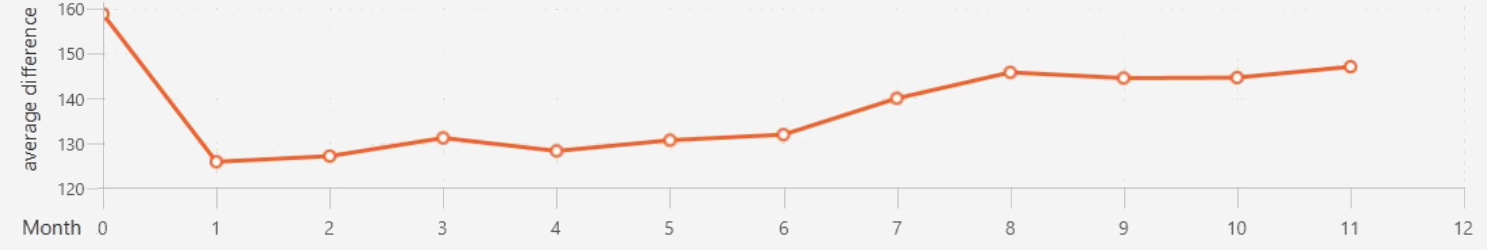} 
\caption{{Difference between the average users reputation, 
$t_a=8*10^4$ days.}}
\label{diff}
\end{subfigure}
\caption{Dynamic reputation graphs for {\it MathOverflow} vs {\it DIBRM}.}
\label{arseny:ta}
\vspace{3mm}
\centering
\begin{subfigure}{0.89\textwidth}
\includegraphics[width=1\linewidth]{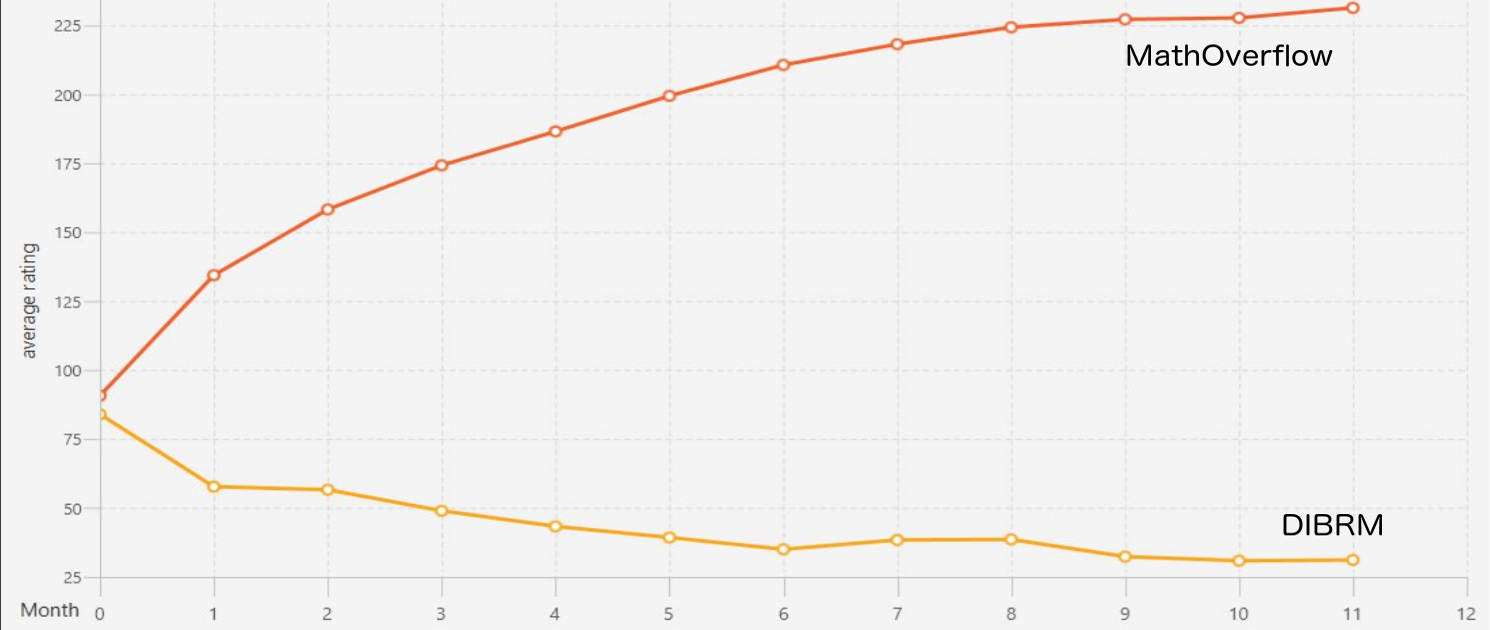}
\caption{{Monthly Average rating change. $\beta=0.9$ was used. $\mu_D$ is lower than $0.6$. The difference grows larger as in  Figure~\ref{initial_ta} ($t_a=1$).}}  
\label{initial_beta}
\end{subfigure}
\begin{subfigure}{0.89\textwidth}
\includegraphics[width=1\linewidth]{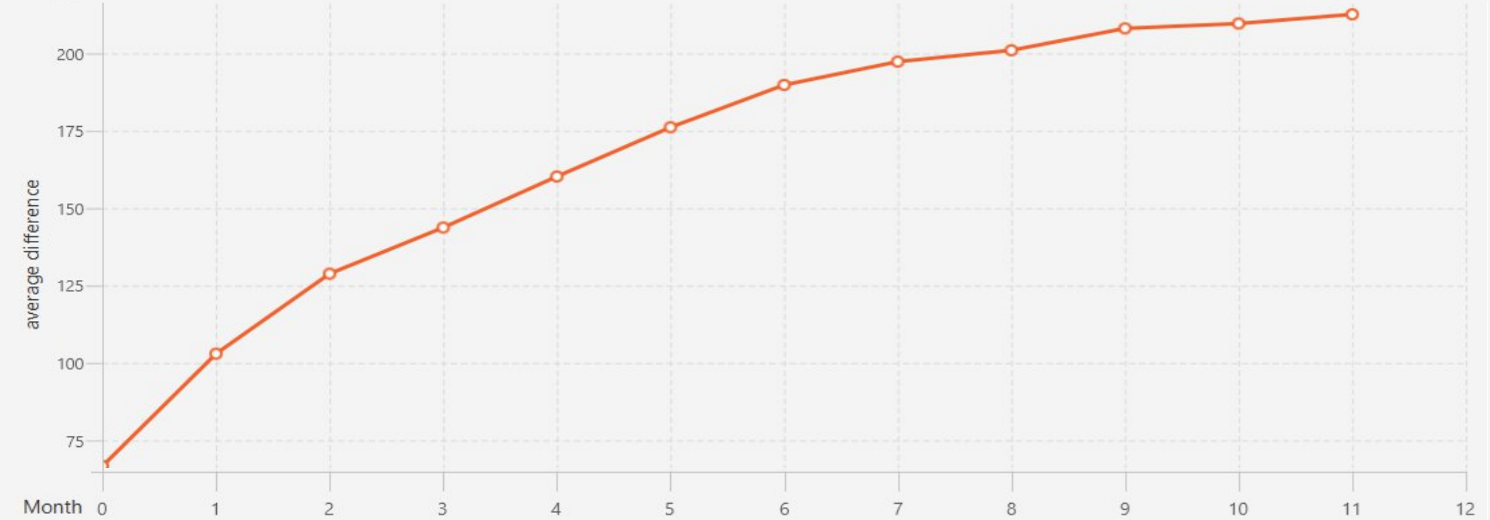} 
\caption{{Monthly Difference between the average reputations  of  users 
with $\beta=0.9$ (similar behavior with Figure~\ref{avediff}).}}
\label{avediffb}
\end{subfigure}
\caption{Dynamic reputation graphs for $\alpha=2$, $t_a=2$ days and $I_b=4$.}
\label{arseny:beta}
\end{figure}

\noindent \textbf{MathOverflow}
An open dataset  by \textit{MathOverflow} was  downloaded\footnote{\url{ https://archive.org/details/stackexchange}}. 
It contains information about votes, comments, posts and users from the launch of the platform, 28 October 2009,  to 12 December, 2017. A portion was selected  including `votes', `comments' and `posts' (from 28 October 2009 to 28 November 2010, 4793 users). One month data was used for measuring the difference between the DIBRM and {\it MathOverflow} models.  
From Figure~\ref{initial_ta}, the estimation by DIBRM differs more over time because of the dynamic properties. While with the {\it MathOverflow} model users tend to increase their reputation over time, in the DIBRM model users tend to lose reputation values due to the $\beta^{\Delta_n}$ factor. This difference decreases with a larger amount of months in the historical DIBRM model.
As in Figure~\ref{ave}, the two lines become parallel.

\newpage

\subsection{Discussion}\label{discussion}
In general, the experiments conducted with {\it MathOverflow} data showed better results when compared to those executed with {\it Reddit} data. In fact, there was no negative $\mu_D$ values and the difference between the DIBRM ratings and the actual rating was smaller with data from {\it MathOverflow}. This is likely due to the size of the dataset:  larger for {\it MathOverflow} containing more temporal information. Unfortunately, with  {\it Reddit} it was not possible to collect a significant amount of data. 
Moreover, the {\it Reddit} rating algorithm is officially not made publicly available, contrarily to {\it MathOverflow}. Results showed how the DIBRM model can approximate other rating systems accurately and with minimal information. Therefore it can be suitable as an inferential system of reputations of online users when available data is limited.

\begin{figure}[h!]
\centering
\begin{subfigure}{0.90\textwidth}
\includegraphics[width=1\linewidth]{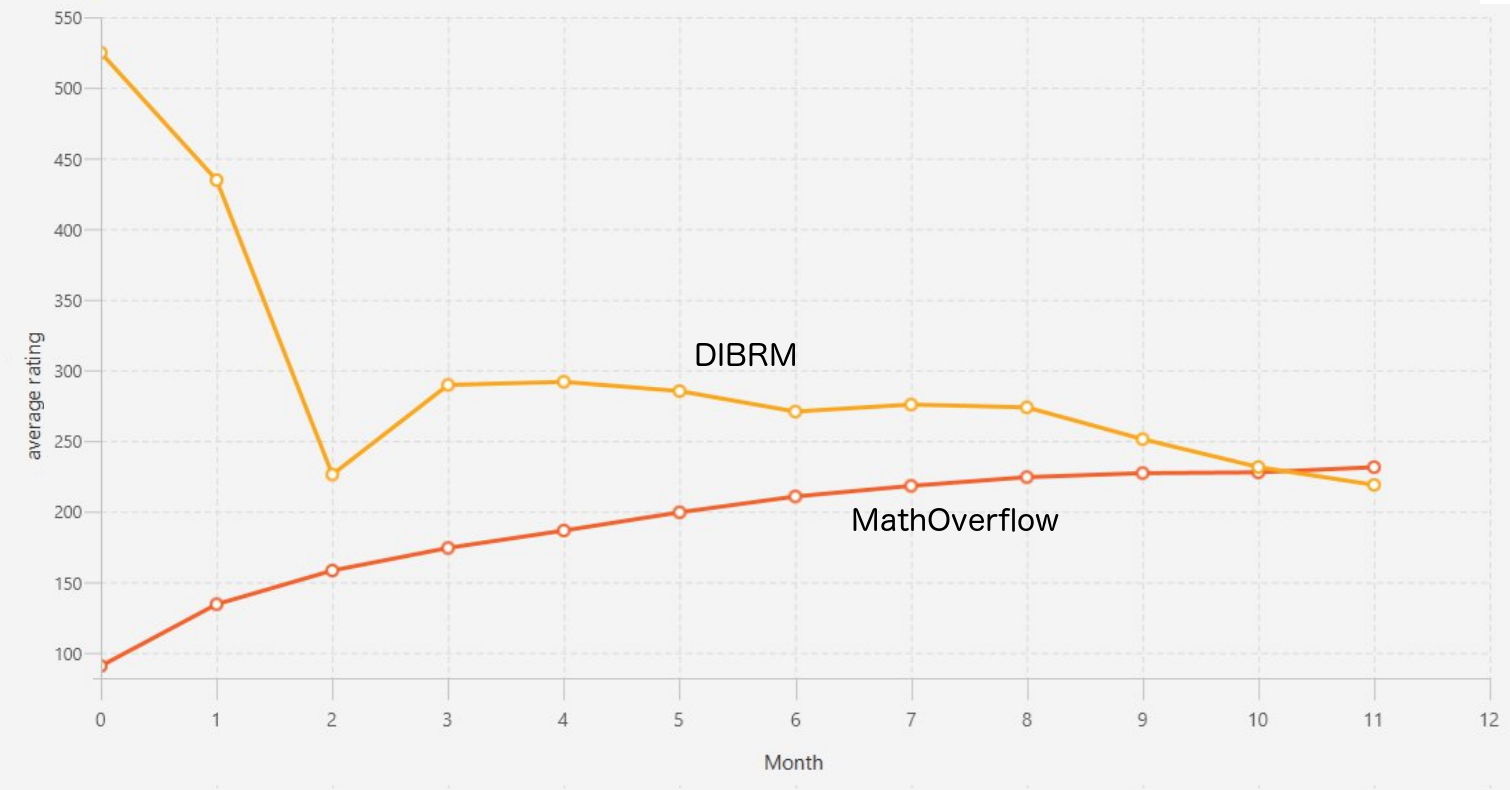} 
\caption{Average rating change over a month. $t_a=2$ days, $\beta= 0.99$ and $I_b=4$. Unlike other cases, difference is decreasing over time.}  
\label{a8}
\end{subfigure}
\begin{subfigure}{0.90\textwidth}
\includegraphics[width=1\linewidth]{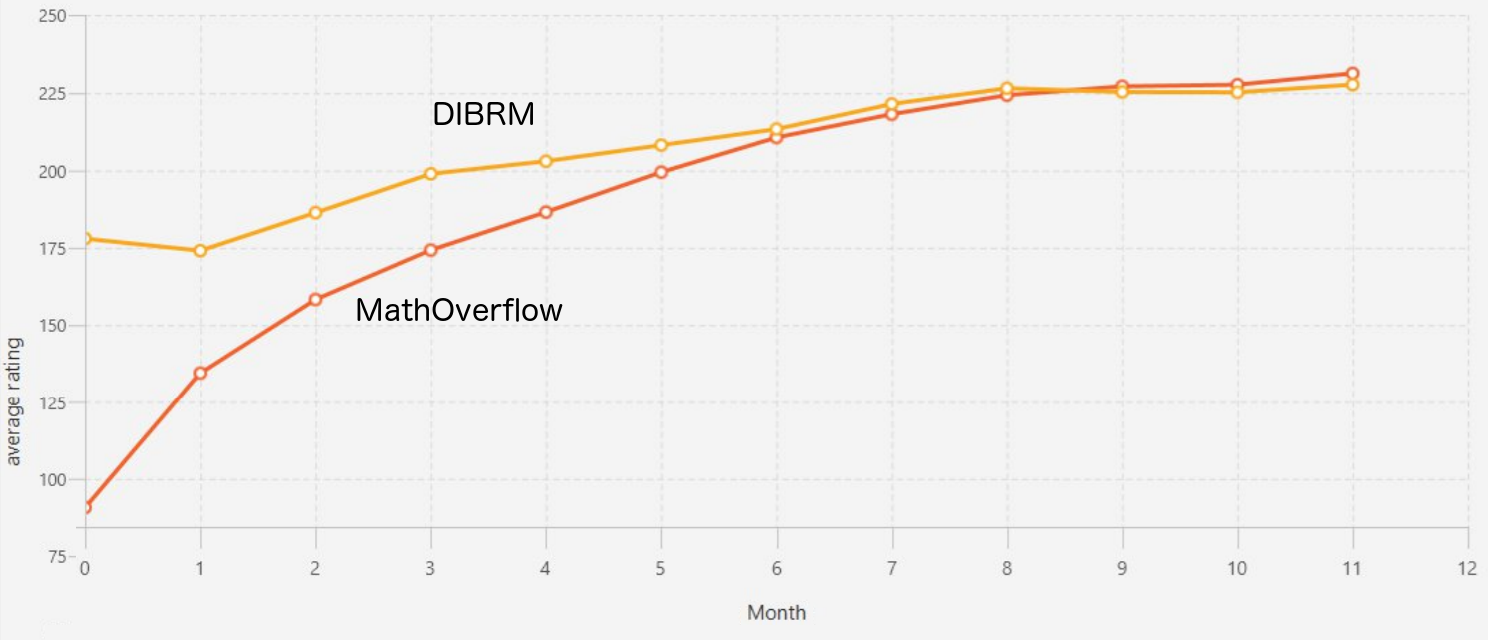} 
\caption{{The best set of parameters that makes {\it DIBRM} and {\it MathOverflow} ratings  closer is: 
$t_a=2$ days, $\beta=1$, $\alpha=1.4$ with approximation, $\mu_D$  $0.95$.}}
\label{best}
\end{subfigure}
\caption{Dynamic reputation graphs for the best scenario.}
\label{arseny:best}
\end{figure}

\section{Conclusion}\label{conclusions}
Several studies have been conducted on reputation models, with such models having a draw back with the anonymity of users, an important concept in modern research.
In this paper, a comparative research has been performed using an existing reputation model, the Dynamic Interaction-based Reputation Model (DIBRM). 
This model expresses trust of users by employing temporal frequency of their interaction in a given online community. In detail, the DIBRM model was applied to other popular social networks, namely {\it Reddit} and {\it MathOverflow}, extending its previous application with {\it StackOverflow}.
Through a set of experiments, it has been observed that DIBRM, without any rating and user profile information, can mimic the behaviors of reputation algorithms used by different online social platforms with a good degree of accuracy. It also shows how its accuracy can be greatly improved when enough interaction data is available. This implies that DIBRM captures the modern essence of reputation mechanisms without accounting for the underlying structure of online platforms and the nature of interactions. Future work can focus on building an automatic tuning mechanism for the parameters required within DIBRM, giving it the ability to become truly versatile.
Additionally,  further empirical validation of DIBRM can be done by running similar comparisons with data from other interaction-based online platforms. 

\bibliographystyle{spmpsci}
\bibliography{main}

\end{document}